# Quantum Coherence in Loopless Superconductive Networks

M. Lucci[1], D. Cassi[2], V. Merlo[1], F. Romeo[3], G. Salina[4], and M. Cirillo[1]

[1]Dipartimento di Fisica and *MINAS Lab*, Università di Roma "Tor Vergata", I-00133 Roma, Italy

[2]Dipartimento di Scienze Matematiche, Fisiche ed Informatiche, Università di Parma, I-43124, Parma, Italy

[3]Dipartimento di Fisica "E. R. Caianiello", Università di Salerno, I-84084 Fisciano (SA), Italy

[4]Istituto Nazionale di Fisica Nucleare, Sezione Roma *Tor Vergata*, I-00133 Roma, Italy

## Abstract

Measurements indicating that planar networks of superconductive islands connected by Josephson junctions display long range quantum coherence are reported. The networks consist of superconducting islands connected by Josephson junctions and have a tree-like topological structure containing no loops. Enhancements of superconductive gap over specific branches of the networks and sharp increases of pair currents are the main signatures of the coherent states and, in order to unambiguously attribute the observed effects to branches being embedded in the networks, comparisons with geometrically equivalent, but isolated, counterparts are reported. Tuning the Josephson coupling energy by an external magnetic field generates increases of the Josephson currents, along the above mentioned specific branches, which follow a functional dependence typical of phase transitions. Results are presented for double comb and star geometry networks and in both cases the observed effects provide positive quantitative evidence of the predictions of existing theoretical models.

# 1) INTRODUCTION

In 1979 it was proposed [1] that a Berezinski-Kosterlitz-Thouless (BKT) phase transition [2] in two-dimensional superconducting films could occur and its possible evidence was linked to the sheet resistance of the films; later, interest raised for possible observation of BKT transition on two-dimensional arrays of superconductive islands shaped to form closed loops when connected by Josephson junctions [3]. The evidence of the BKT transition of the arrays would be the activation, below a given temperature, of vortex-antivortex pairs and the resistive transitions accompanying their motion. The topic had noticeable developments in the early 80s, but the 1986 discovery of High Tc Superconductors (HTCS) [4] dragged most of the attention of the condensed matter community for understanding the properties of the cuprates. It was so that even the investigations of BKT and topological transitions in superconductors were somewhat asided by the wave of high Tc superconductivity research and perspectives. It is worth recalling that a BKT transition is a phenomenon predictable in superconductive systems only when the temperature of these stands safely below the superconducting transition temperatures. It is then understandable that exceptional rises of these temperatures (up to *120 K* at atmospheric pressure) of the new materials would primarily attract research work.

The development of laser cooling techniques in mid 80s [5] and the consequent discovery of Bose-Einstein Condensation (BEC) in alkali atoms vapours in mid 90s [6] raised a noticeable burst of activity related to the physics of systems described by macroscopic wave-functions and, within this framework, attention for BKT phenomenology and topological phase transitions in condensed matter grew again [7]. Attempts to combine BEC physics and networks connectivity have been provided, both from a theoretical physics [8] and mathematical [9,10] point of view. These efforts, analyzing populations of bosons distributed over the sites of discrete reticles having specific nearest neighbour connectivity (and adequate intra-sites coupling potentials), demonstrated that in these systems peculiar long range macroscopic wavefunction configuration and thermodynamic features typical of BEC transitions, are possible. Although the theoretical predictions were first conceived for reticles of bosons of optical networks, attention was also dedicated [11] to the possibility that the same effects could be observed in arrays of superconducting islands connected by Josephson junctions containing no superconductive loops: in this case the role of the bosons on the reticles sites should be played by Cooper pairs. A strong argument in favor of this intuition comes from noting that the Bose-Hubbard model, under macroscopic occupation of the lattice sites, implies a non-linear Schrodinger equation (NSE) for the condensate wavefunction. The NSE, in turn, can be organized in the form of a generalized Feynman's model [12] describing the Josephson coupling among the lattice sites (superconducting islands). The aforementioned observation suggests a close analogy between hopping of bosons in optical networks and Cooper pairs dynamics on networks of superconducting islands coupled by the Josephson effect.

From the experimental point of view, to our knowledge, there has not been much activity dedicated to investigate experimentally the bosonic nature of Cooper pairs and therefore a probe of the predictions of ref. 11, relying on Cooper pairs acting as hopping bosons, would costitute evidence of such an integer spin behavior. Beside the exposed analysis in terms of BEC it is worth noting that a parallel between a BKT transition and the phenomena predicted in ref. 11 exists: the "hopping" of bosons between all the islands of the superconductive arrays can generate long range correlation of the wave functions in the system only when the Josephson coupling energy is of the order or less than the thermal energy [11]. This is analogous to say, in BKT language, that a long range correlation leading to vortex-antivortex dissociation can exist only below a critical temperature.

The systems we herein investigate present a two fold topological aspect: the first concerns the spatial long range coherence of macroscopic wavefunctions, a phenomenon nowadays addressed as "topological order". The second comes from the fact that the "topological order" in the specific case can be achieved through specific node-to-node connections, which are the basis for the topological analyses of electrical networks and graph theory. We will show that evidence of long range order in our systems comes both from Josephson supercurrent peculiarities and from gap increases of the islands of the discrete systems we investigate. It is herein shown, by new samples, new data, and more quantitative arguments, that previously observed effects on double-comb [13,14] and star graph arrays of Josephson junctions [15,16] can be safely attributed to peculiar topology (in terms of connectivity) of these structures.

## 2) DOUBLE COMB-LIKE STRUCTURES

In Fig. 1 CAD designs of the superconducting networks whose features are investigated in our experiments is displayed. Fig. 1a shows a portion of the backbone of a double comb planar array in which the "fingers" of the comb are just superconducting shorts. We call backbone just the linear array of islands from which the fingers depart. The arrows indicate the location of Josephson junctions (JJ). In Fig. 1b instead we can see a double comb structure having the fingers made of superconductive islands connected by Josephson junctions: in this structure fingers are departing above and below the backbone so that each island of the backbone is connected to *4* neighbour islands while each island of the fingers is connected only to two. All the superconductive islands have been designed to have in the final chip the same volume, which was possible due to the different thicknesses of base and contact electrodes layers. Responsible for the connections between the islands are the Josephson junctions which are the (*3μm x 3μm*) squares indicated by the arrows. The superconductive islands are made in niobium and the Josephson tunnel junctions are fabricated in the *Nb-NbAlOx-Nb* technology [17]; the backbone branch was formed by *101* superconductive islands while each finger branch has *50* islands. When measuring current-voltage (*IV*) characteristics of fingers, however, we bias two aligned fingers in series making a connection of *101* islands as well (the extra island of two aligned fingers is a backbone island connecting these).

In Fig. 1c we show what we call a "reference" backbone array, namely backbone array with no fingers connections attached to it. It is worth noting that the superconducting "shorts" generating the fingers in Fig. 1a are extentions, in alternate succession, of base or contact electrodes of the backbone islands and have all the same volume. Both reference and the "shorted-fingers" arrays were designed to perform comparative analyses with the current-voltage characteristics of the backbone array embedded in a graph structure (Fig. 2b) : the junctions connecting the backbone islands to the fingers are deliberately missing in Fig. 1a and Fig. 1c in order to isolate geometrical from connectivity effects.

The *IV* characteristics we investigate are those obtained biasing the arrays from four final contact pads (two to the left end and two to the right end) and we observe the voltage sum of *100* junctions, in a typical four probes configuration. In order to isolate the arrays from the final contact pads we insert normal metal layers between array's ends and contact pads. In general, for the present design, *N+1* superconductive islands (including the two final "contact" island) generate *N* junctions. The data we present herein were all obtained with the samples immersed in a $^4He$ bath. The helium dewar was put inside a $\mu$-metal cylinder and an additional cryoperm shield was surronding the samples in the bath at *4.2K*. The current noise level for our combined analog/digital acquisition system is of the order of few nanoamperes; the voltage signals generated at the "cold" end by the chips are amplified at room temperature and fed to a data acquisition system. The "philosophy" behind our chip design is, first, to identify the effects generated on the *IVs* of the backbone array by the fingers eliminating these (as in Fig. 1b) and also investigate what is the effect of fingers just made of short circuits, i.e. not including Josephson junctions-connected islands (see Fig. 1c).

A comparison between the three arrays of Fig. 1 is shown in Fig. 2. In Fig. 2a we see that the current-voltage characteristic of a backbone array has higher Josephson currents and higher gap-sum voltage with respect to the reference array, a result consistent with those reported in previous papers [13,14]. In this figure, however, we also show the current-voltage characteristics of the backbone array with superconducting shorts of Fig. 1c: we can see that for this array the Josephson currents are, up to *100mV*, equal to the currents of the backbone array and after this voltage the "shorted fingers" array has higher Josephson currents. Note that the sort of "grass" appearing on the switching distributions of the samples does not depend on noise of the measurement apparatus, but it is an intrinsic characteristic of the junctions of the arrays having slight barrier/geometrical defects generating quasi-particle current dispersion which add to the Josephson currents when measuring the series connection.

It is important to point out that when tracing current-voltage characteristics, as we did for Fig. 2a, the top of the Josephson currents (just before the switches from zero voltage occur) do correspond to states in which the Josephson potential, coupling the current-biased junctions, becomes very low [19], while, for current unbiased junctions, lying on the fingers of the arrays, the Josephson energy is much higher than the energy of thermal fluctuations at *4.2K* (roughly a factor *30* for a maximum Josephson current of *5.6μA*). In these conditions we might not fully exploit the characteristics of the phase transition predicted in ref. 11, but the observed effects show that: *i)* the fingers in a double comb structure are necessary to observe the increase of the Josephson currents , *ii)* the increase we observe is, over roughly *15%* of the junctions of the backbone array, of the same order of what one could observe if the fingers arrays were to be replaced by a single superconducting short. In other terms, *i)* and *ii)* provide evidence that the effect of the fingers branches is equivalent to that of a single superconducting film (described by a single wavefunction) as far as the amplitude of Josephson currents are concerned. We suppose that above *100 mV* the backbone array currents of the complete comb (Fig. 1b) are smaller than those of the shorted-fingers array (Fig. 1a) since, as discussed above, the Josephson energy for the unbiased junctions of the fingers is too high with respect to thermal fluctuations that can generate hopping and charge migration between islands. For reasons that will be clear few paragraphs below, we believe that is a rather reasonable explanation for the difference.

In Fig. 2b we have plotted the normalized distribution of the gap amplitudes of all the junctions of backbone and reference array of one chip: we see that the distributions are mostly gaussian (the curve through the data) and that the peak of the backbone array is *3%* ahead of the peak of the reference array. The information coming from these distributions suggests that the junctions of the backbone array have a gap increase distributed around a mean value and this is a solid evidence of a regular increase of superconducting gap (and condensation energy) all over the islands of the array. The average value of the distributions returned by the fits are respectively *(2.687±0.005)mV* for the backbone array and *(2.599±0.003)mV* for the reference, where the error is relative to the standard error of the mean $\bar{\sigma} = \sigma/\sqrt{N}$ where $\sigma$ is the standard deviation to be associated with each measurement and *N* the number of gap measured *(100)*. We note that the statistical errors are at least one order of magnitude larger than the instrumental errors which range in the tens or few hundreds nanovolt order of magnitude. The plot of Fig. 2b was obtained grouping the *100* gap readings in intervals of *20μV*. A *Student unpaired t-test* over all the data (*100* for each curve) relative to the gaussian curves in Fig. 2b gave as result that the probability of the difference of the means is being caused by statistical fluctuations is less than *$10^{-4}$* .

In terms of gap dependence on temperature [18], the observed gap difference would correspond, at *4.2K*, to a temperature difference of *0.8 K* between the two arrays. The backbone array then should be "colder" than the reference array of *0.8 K*: this effect is not be attributed to "fingers" cooling (or else) since the gaps of backbone with the shorted fingers (Fig. 1a) have the same values of the reference array. It is also worth recalling that we are measuring in a liquid helium bath and such thermal gradients are not really conceivable for samples dissipating few microwatt of power and are less than a millimeter apart on the same chip (latent heat of evaporation of *$He^4$*, at atmospheric pressure, is *21 kJ/kg*). Thus, the increased gaps of the junctions on the backbone has to be uniquely attributed to the specific topological (in terms of connectivity) properties of the double comb structure. Fig. 2b shows that the statistical distribution of the gaps has a gaussian shape just like that of the geometrically equivalent array meaning that, apart for the statistical fluctuations, a uniform increase of the gap is distributed all over the backbone islands. It must be noted that in Fig. 2b the distribution of the data (collected using the same "voltage bins" for the statisical analysis) is wider for the backbone array meaning that the gap increase on the backbone has a higher disuniformity .We will show in the next paragraph that this dispersion depends on finite size effects on the comb for which the gap increase tends to be more reduced toward the end of the structure.

In Fig. 3a we show the gap increase of the central finger, CF, (meaning by this the series connection of two aligned fingers) located in the center of the comb structure (red) with respect to the very last finger LF, (black). Here we can see that the central finger array has a gap-sum voltage higher than the one of the last finger; in Fig. 3b, were we show the statistical distribution of the gaps along central and last finger we can see the difference between the central peaks of reference and lateral array. In particular, the mean gap value of the central finger is *(2.675±0.004) mV* while the mean of the last finger is *(2.641±0.008) mV*. Here again the plot was obtained by grouping the data in intervals of *30µV* and a *Student unpaired t-test* performed on all the available gaps (*100* for each array) provides a rather high confidence that the observed differences between the two means cannot be generated by statistical fluctuations (the contrary has a probability less than *$10^{-4}$*). Physically the two distributions tell us that there is a difference as we move along the backbone toward the physical ends of the double comb structure and therefore it is reasonable to expect, as we mentioned before, a wider distribution of an embedded backbone array with respect to his geometrically equivalent as we see in Fig. 2b. In Fig. 3b we see that, beside having the central peak for a higher value of the voltage, the distribution of the central finger array is even more peaked than that of the lateral array, meaning that there is more "coherence" (in terms of uniform increase) of the gap values" along the central finger which is deeply embedded in the double comb structure.

Hereafter we present a comparison between the magnetic field behavior of the arrays of Fig. 1 by applying an external field. The magnetic field, generated by a superconducting niobium solenoid, has a direction lying in the planes of the arrays and perpendicular to the backbone arrays lines, i.e. parallel to the fingers. The field depresses uniformly all Josephson critical currents $I_c$ of the junctions of the arrays, and related coupling energies $\Phi_0 I_c/2\pi$, through the Faunhofer pattern and therefore any thermal effect due the flipping of Cooper pairs between the islands can be enhanced. In Fig. 4a, we show the dependence of the difference *ΔI* between average Josephson currents of backbone array and reference array averaged in the interval *(100-150) mV*. The difference is normalized to the value of the average reference current at each specific magnetic field value and therefore what we measure is the relative percentage of the increase in Josephson current of the backbone array $\frac{\Delta I}{I}(B) = \frac{const}{\sqrt{B_c - B}}$ .

We see that, when the magnetic field is around *27.5 G* a sharp increase of the percentage takes over (see inset where the abrupt increase is evident) up to the point that the average currents of the backbone become more than *3* times that of the reference array. The curve we just wrote is a typical phase-transition dependency describing the experimental points with a coefficient of determination ($R^2$) different from the unity only for *4* parts over *$10^4$* ; the experimental error bars are essentially uninfluential for the fit since all the experimental data are intercepted by the theoretical curve. In any case, the maximum uncertainty to be associated to the points of the plot is of the order of *15%* (due to the averaging of the values). Also note that right after the maximum the data attain zero value because we are not far from the first minimum of the Josephson current which occurs for *B=30 G* and therefore the static vortex trapped in each junction changes substantially the static distribution of the phase and sets Josephson currents to zero.

In Fig. 4, we also show the results obtained now measuring the magnetic field response of the shorted-finger array. In this case, we can see a much more limited increase of the Josephson currents of the short-finger array backbone with respect to the reference backbone array. The latter observation implies that the sharp increase we observe in Fig. 4 is generated by the granularity of the finger array, made of coupled superconductive islands. In zero applied field, for a maximum Josephson current of *5.6 $\mu$A*, the Josephson coupling energy is roughly *30* times higher than the thermal energy at *4.2K* *($5.8 \times 10^{-23}$ J)*. However, at the field of *27 G* the average Josephson current of the reference array is *0.270$\mu$A* which leads to a Josephson zero-bias energy of *($8.9 \times 10^{-23}$) J* , a bit above the thermal energy written before and therefore thermal hopping is possible all over the islands of the comb array causing a sharp increase of the Josephson currents of the backbone. Thus, the behavior under the applied magnetic field of the three arrays show, unambiguously, that the observed current increases of the backbone array embedded in the double comb structure is due to a macroscopic phase transition in the sense exposed in ref. 11: when the Josephson coupling energy between the islands becomes of the order of the thermal energy the migration of pairs from the fingers toward the islands of the backbone increases sharply, leading to a noticeable increase of the Josephson currents. On the fingers of the shorted fingers arrays we do not have the gradient of charge carriers generated by Josephson energy modulation and there can be only very slight changes varying the magnetic field which are likely due to the depression of the energy of the two junctions of the backbone islands. We note that this comparison between the "shorted-fingers" backbone array and the backbone array embedded in the whole comb structure is a fundamental completion of the analysis reported in ref. 14. It is established now that the noticeable increses of the currents when the Josephson energy becomes of the order of the thermal energy is due to the "granularity" of the fingers.

# 3) ANALYSIS OF THE EXCESS VOLTAGES IN STAR ARRAYS

It is worth noting that, while the amplitude of the Josephson currents of the backbone increases noticeably with respect to those of the reference array, as shown in Fig. 4, the increased value of the gap energy does not depend on the external magnetic field. While the predictions of ref. 11 were dealing with bosons coupling through adequate potentials, the value of the gap energy is strictly related to the condensation energy of the superconductors and no specific predictions were made. An increased gap in our backbone structure implies an increased superconducting transition temperature, as clearly shown in ref. 13. This phenomenon, however, is not directly related to the topological BEC described in ref. 11: as we said earlier, a BEC approach of these authors relies on an existing superconducting condensate over the array, just like a BKT relies on the existence of a superfluid for vortex-antivortex dissociation. Also note that the relative increase of the Josephson junctions of the backbone array with respect to its geometrically equivalent reference is substantially higher than the relative increases of the gaps, differently from what one would expect from Ambegaokar-Baratoff prediction [17,19]; this statement can be clearly appreciated in Fig. 3a where we see that the Josephson currents increase goes up to *12%* while the increase of the gaps is of the order of *3%*. This phenomenon was already evident in the temperature-dependent characterizations reported in Fig.2 of ref. 13 where one could clearly see the gap sum of the backbone array being slightly higher than the gap sum reference array, but that decreasing the temperature down to *1.2K*, the increase of Josephson currents became noticeable (*15%*) while the gap difference remained the same: being the samples in that experiment immersed in superfluid helium, the increase in Josephson current increases could hardly be attributed to thermal effects and gradients.

Motivated by the arguments described in the previous paragraph we decided to investigate systematically the phenomenon of increase of superconducting energy gap and transition temperature in graph, tree-like arrays. A recent publication, employing an approach based on the De Gennes-Alexander model [21] for granular superconductors has demonstrated that the superconducting transition temperature can be amplified in systems with specific connectivities and, in particular, star-shaped arrays of junctions [21]. In order to test the theoretical predictions of this model, we have fabricated specific samples consisting of strar-graph arrays with different numbers of rays and the CAD design of a sample with 18 rays is shown in Fig. 5a (top). According to the theory, the superconducting transition temperature $\mathcal{T}_c$ of a star-shaped array with $p$ rays can be written as (see Eq. 27 of ref. 21):

$$\mathcal{T}_c = T_c \left[ 1 + \frac{\mathcal{D}}{\alpha T_c} \left( \frac{p}{\sqrt{p-1}} - 2 \right) \right], \qquad (1)$$

being the latter temperature amplified compared to the critical temperature $T_c$ of a single disconnected island. Interestingly, no amplification can be observed for $p = 2$, being the latter a condition topologically equivalent to a linear array of coupled islands (like in Fig. 5a, bottom). Equation (1), depending on the coupling energy $\mathcal{D}$ and on the parameter $\alpha$ of the Ginzburg-Landau theory, suggests that the superconducting transition temperature of a star-shaped array can be enhanced by increasing the number $p$ of rays. Moreover, according to the BCS theory, the superonducting order parameters are related by the formula:

$$\Delta_{star} - \Delta_i \propto \left( \frac{p}{\sqrt{p-1}} - 2 \right), \qquad (2)$$

with $\Delta_{star}$ and $\Delta_i$ the superconducting gap of the array and of the isolated island, respectively.

We have designed star graphs having respectively 8, 12, and 18 rays (this latter shown in Fig. 5a, top of figure). Each star array (103 islands and 102 junctions) had its geometrically equivalent aligned couple of rays as shown in the bottom part of Fig. 5a . In Fig. 5b we show typical *IV* curves showing the gap-sum increase of an embedded star array, an effect better evidenced in the inset where we magnify the gap region for a star array made of *18* rays with respect to its "reference" array having only *2* rays : in this case the relative increase of the array embedded in the star graph structure, shown in the inset, is of the order of *1.35 mV* at *5 μA* . In Fig. 5c we have a plot of the increase of the gap voltage-sum for star arrays having different branches, all measured at a current of $5\mu A$. The increases are all measured with respect to geometrically equivalent samples having $N=2$ and therefore for $N=2$ we have $\Delta V=0$. Each experimental point in the plot refer to a set of four measurements performed in different days and/or different samples and the error bars are obtained from the observed variations of the values.

In Fig. 5c, a comparison between experimental values and theoretical expectations is also presented. In particular, it is expected that the gap voltage-sum is described by the relation

$$\Delta V = \mathcal{A}\left(\frac{p}{\sqrt{p-1}} - 2\right), \tag{3}$$

which is a direct consequence of Eq. (2). The full-line curve in Fig. 5c is obtained by using Eq. (3) with the best fit parameter $\mathcal{A} \approx 0.56\ mV$. From the way the theoretical line fits our results we conclude that the gap increase in the star graphs is consistent with a "granular" model of superconductors. As far as star graphs arrays are concerned, we must specify that even the effects described in previous publications [14,15] for the increase of the Josephson currents have been observed in the present experiments. The aforementioned effect, namely the increase of the Josephson currents of several junctions, also evident in Fig. 5b, was roughly the same for stars with different numbers of rays (at least the number of rays we investigated). However, it is worth pointing out that increases of Josephson currents like those shown in Fig. 4 were reported in a previous publication even for star arrays [16].

## 4) CONCLUSIONS

Our results on graph-shaped, loopless networks of coupled superconducting islands show clear evidences of collective behavior in these systems. Some effects appear to be specific of a BEC topological condensation favored by the hopping of pairs (seen as bosons) through Josephson junctions and this phenomenon requires the existence of a condensate. However, the superconducting condensate itself, as seen from the variations of the gap energy, is also affected by the specific network topology. We have shown that the two phenomena (noticeable increases of Josephson currents and increases of the gap voltages) are not trivially related. In the star topologies, an increasing number of rays generates an increasing excess voltage which can be well fitted by the De Gennes-Alexander theory for granular superconductors, however, the increase of the number of rays does not generate a consequent increase of the Josephson currents. It also has to be noted that the noticeable Josephson current increases observed in double comb-topology structures do not seem quantitatively relatable to the more modest excess voltages that remain constant under magnetic field and temperature variations. In conclusion we can say that, while the Josephson currents of our structures (the pair current between the islands) is strongly influenced by magnetic field, and temperature, the gap excess seem to be strictly determined by topology of the network and does not show significant changes as a function of field and temperature. Overall, our impression is that we have scratched the top of a reservoir of more intriguing phenomena involving macroscopic quantum condensates.

**Figure captions**

Fig. 1: a) a zoom of the backbone region of a double comb-like structure having long superconducting shorts (thin films) as fingers. In b) and c) we show respectively portions of the backbone region of a complete double comb array, with superconductive islands connected by Josephson junctions on the fingers, and of a reference backbone array, with no fingers. This latter has the same geometrical structure of the backbone arrays in a) and b). The arrows in the figures indicate locations of Josephson junctions.

Fig. 2 : a) Comparison of the current-voltage characteristic of the backbone array (red) of a complete comb structure with that of a "backbone reference" array (black) and "shorted-fingers array" (blue) : we can see that the reference array has both Josephson currents and gap-sum voltage less than those of the backbone array while the shorted finger has more than half of the Josephson currents higher than those of the backbone but the same gap-sum of the reference array; b) statistical distribution of the superconducting gaps of the *100* junctions for reference (BBR) and backbone (BB) array. From the plot we extract that a single gap voltage of an island of a backbone array is *3%* higher than the reference backbone array.

Fig. 3: a) The current-voltage characteristics of central finger (red) and last finger (black); b) the statistical distributions of the gaps of lateral finger (LF) and central finger (CF). We can clearly see the finite size effect of the comb structure through the gap increase of the central finger with respect to the lateral finger.

Fig. 4: The red squares show the relative average increase ($\Delta I$) of the Josephson currents of the backbone of the complete comb (Fig. 1b) with respect to that of the reference array (Fig. 1c) measured for increasing values of an applied external magnetic field. The relative increase is normalized to average current of the reference array for each field: we can see that for a field value of *B =27.5G* the average Josephson current of the of the backbone array is more than three times

higher than the currents of the reference array (as visible in the inset). The black circles show the dependence of the difference between the average heights of reference and shorted-fingers comb and reference backbone arrays: here we can see that there are no sharp increases like for the black squares. This effect is also visually illustrated in the insert where we see that the "shorted fingers" array currents remain much lower than the values attained by the backbone of the complete comb structure.

Fig. 5: a) Top : area of the central island of superconducting star-shaped arrays having *18* rays; bottom: the geometrical equivalent of two aligned rays of the star; b) the current-voltage characteristic of two aligned rays of a star array compared with the geometrical equivalent. The inset shows a magnification of the gap region of a typical sample where S3 indicated is the star embedded two rays array and R3 the geometrical equivalent ; c) excess voltage of gap sum for arrays embedded in stars with increasing number of rays. The excess voltage is measured with respect to the gap sum of two aligned "reference" rays, see a), bottom figure. The dependence in very good agreement with the theoretical curve obtained from a De Gennes-Alexander approach for granular superconudutors (continuous curve).

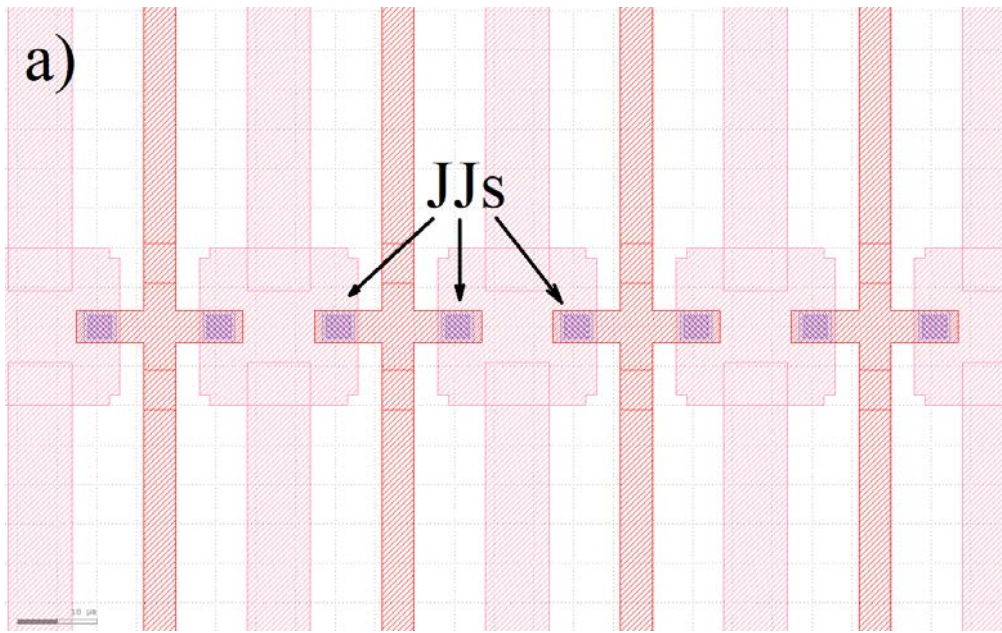

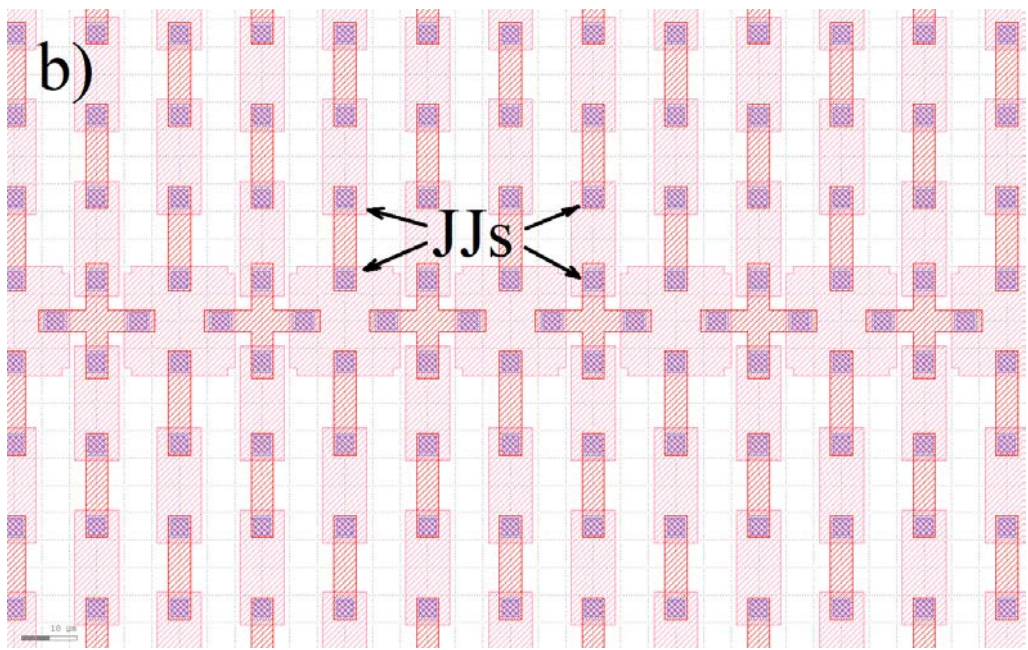

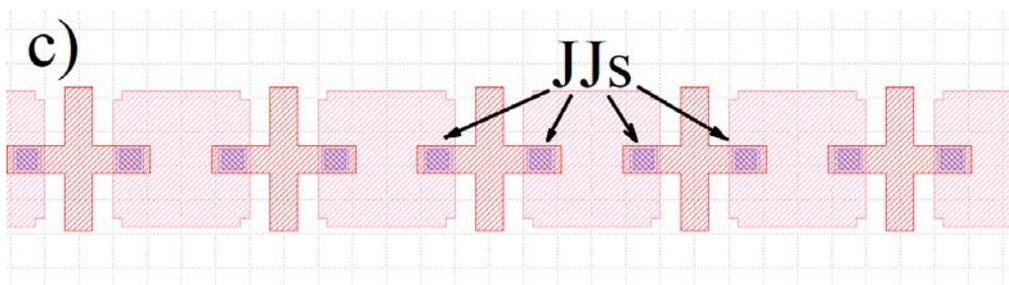

Figure 1

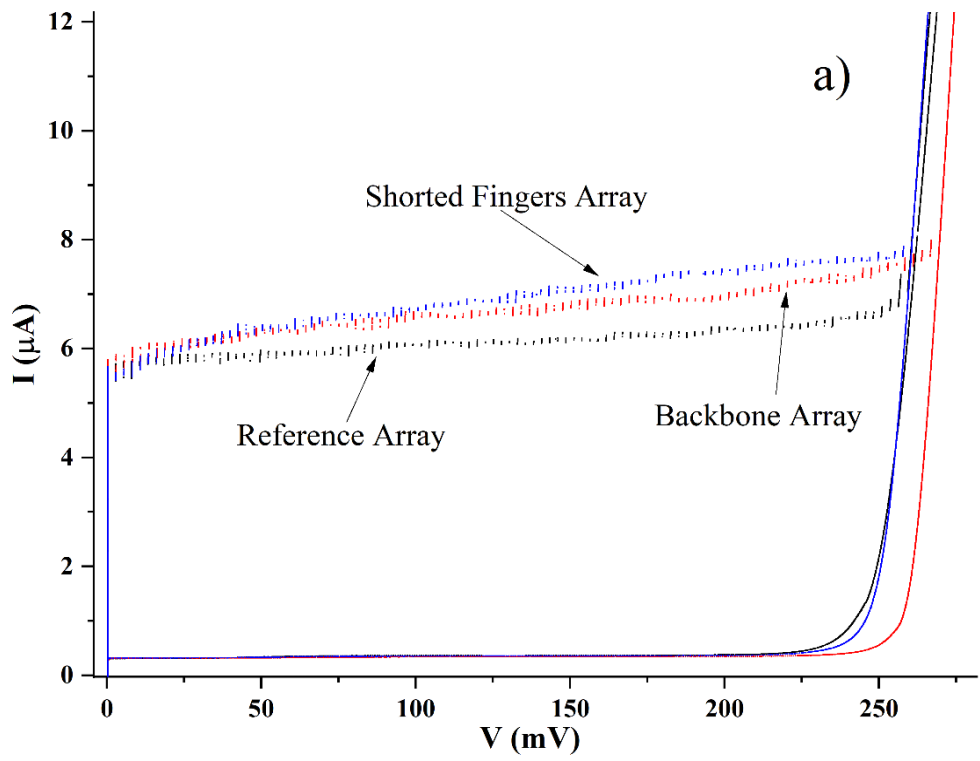

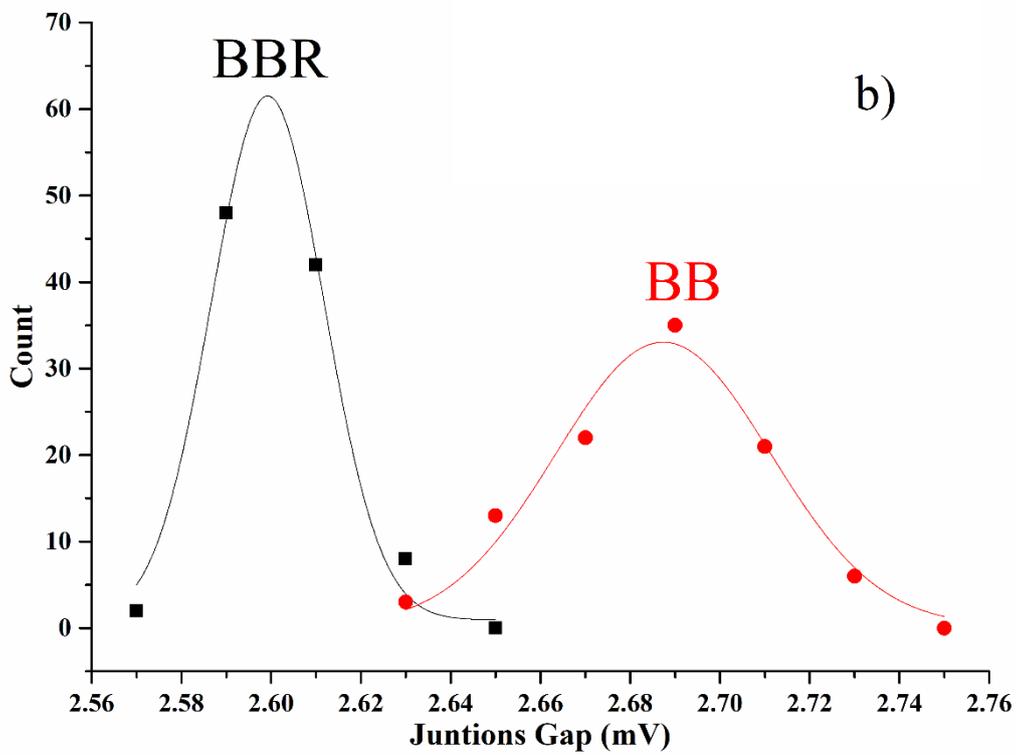

Figure 2

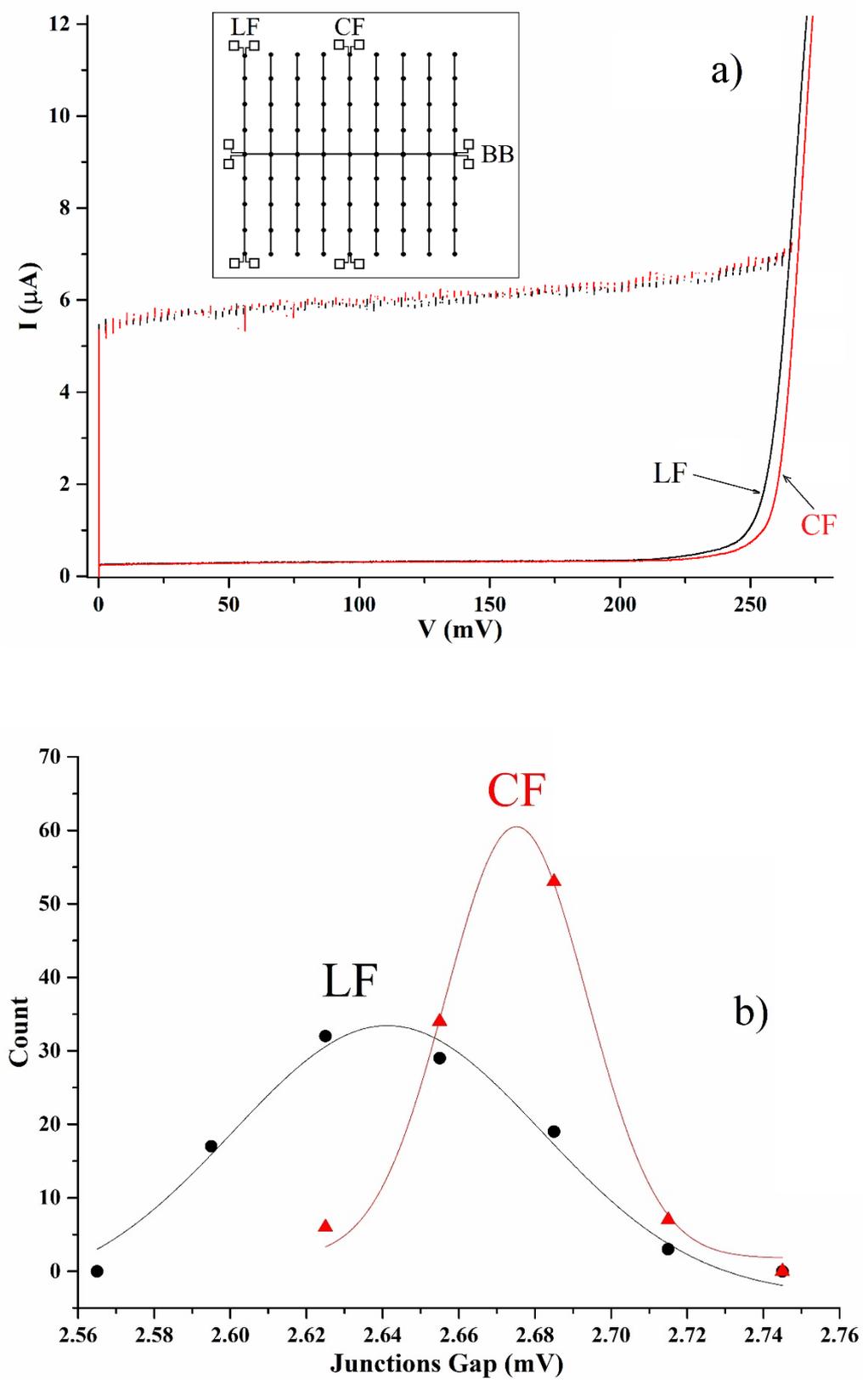

Figure 3

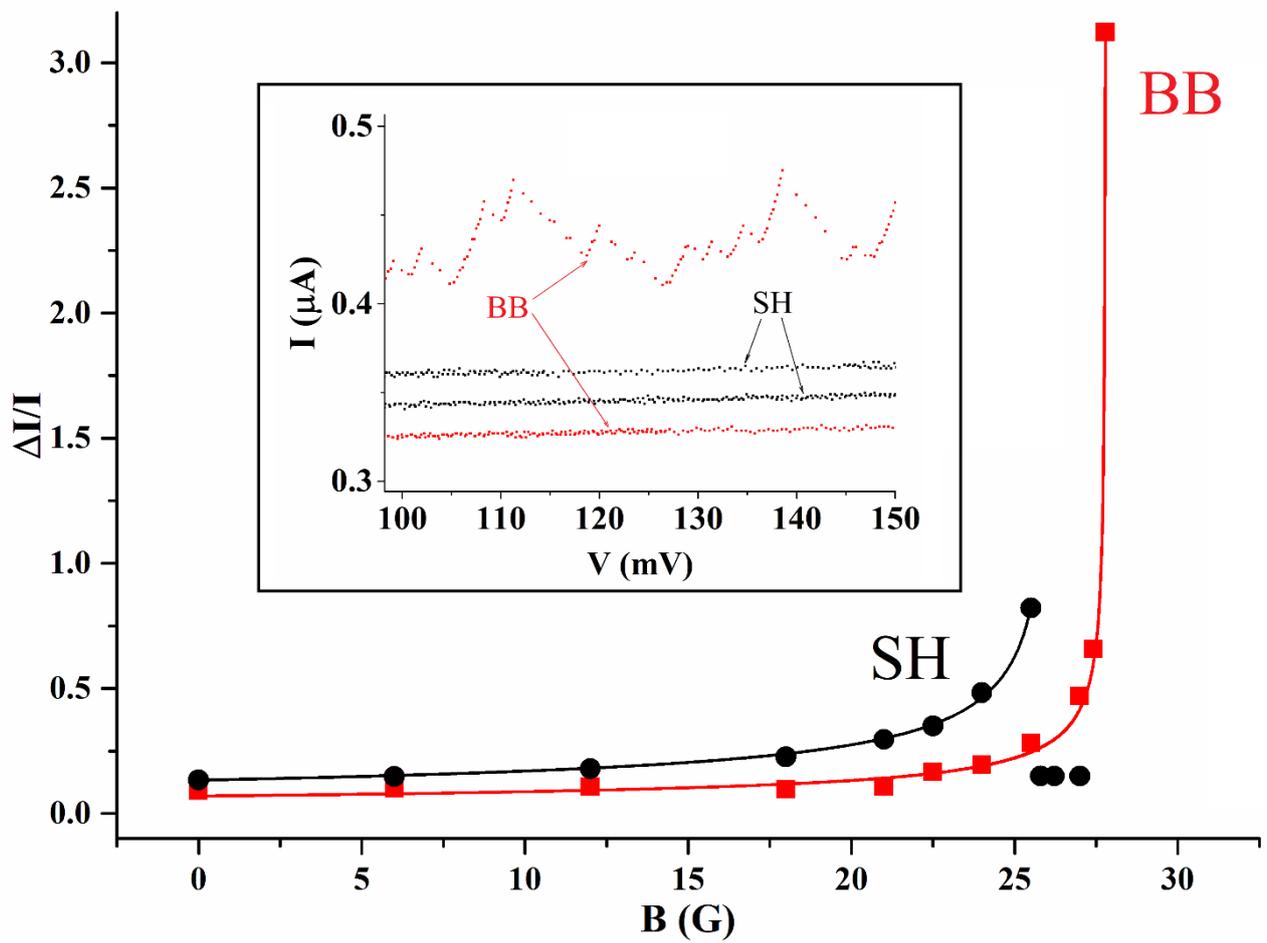

Figure 4

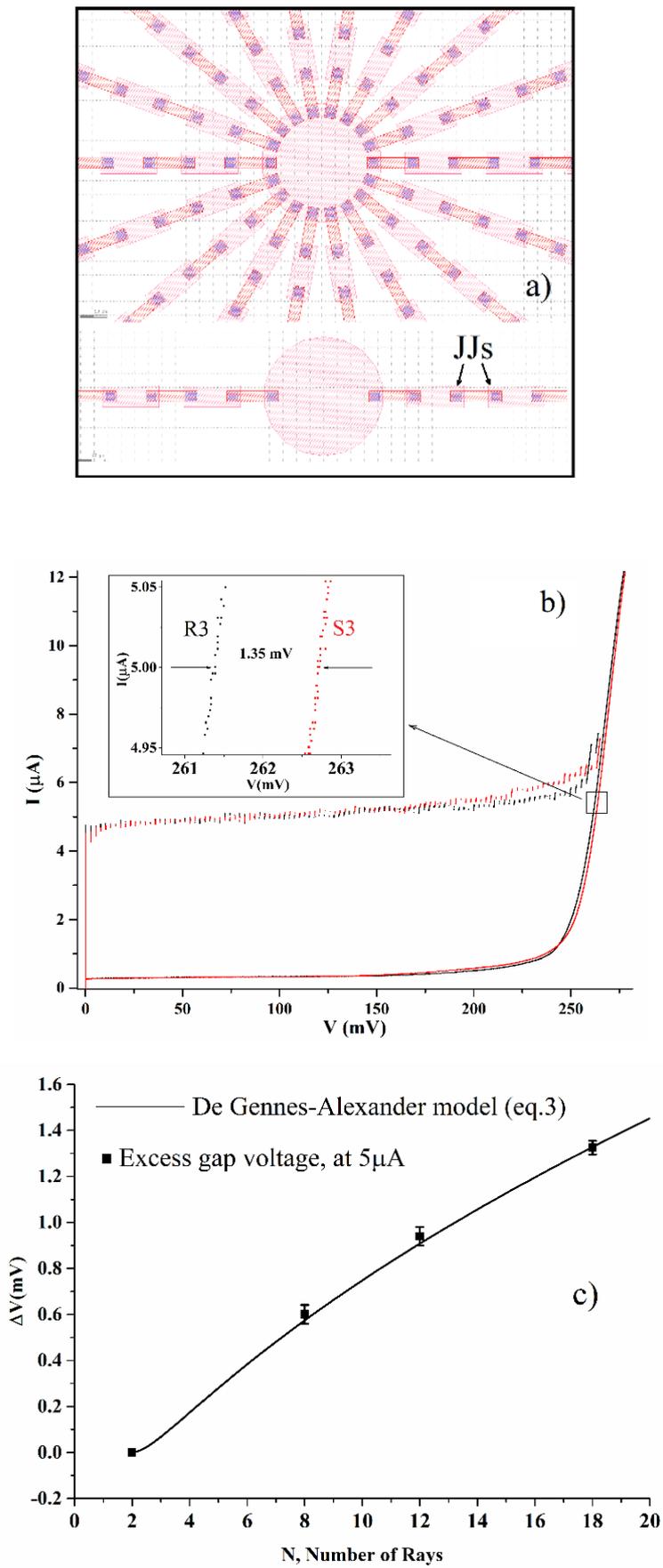

Figure 5